\documentclass[aps,prb,twocolumn,superscriptaddress,noshowpacs]{revtex4-2}
\usepackage[utf8]{inputenc}
\usepackage[T1]{fontenc}
\usepackage{graphicx}
\usepackage{xcolor}
\usepackage{physics}
\usepackage{bm}
\usepackage{amsfonts}
\usepackage{hyperref}

\begin{document}

\title{Mode-locking instability and multiple soliton formation in GaN polariton waveguide cavities}


\author{O. Bahrova}
\altaffiliation{These authors have contributed equally to the present work.}
\affiliation{Institut Pascal, PHOTON-N2, Universit\'e Clermont Auvergne, CNRS, Clermont INP,  F-63000 Clermont-Ferrand, France}
\affiliation{B. Verkin Institute for Low Temperature Physics and
Engineering of the National Academy of Sciences of Ukraine, 47
Nauky Ave., Kharkiv 61103, Ukraine}

\author{V. Develay}
\altaffiliation{These authors have contributed equally to the present work.}
\affiliation{Laboratoire Charles Coulomb, Université de Montpellier, CNRS, 34095 Montpellier, France}


\author{H.~Souissi}
\affiliation{Centre de Nanosciences et de Nanotechnologies, CNRS, Universit\'e Paris-Saclay, France}

\author{C. Brimont}
\affiliation{Laboratoire Charles Coulomb, Université de Montpellier, CNRS, 34095 Montpellier, France}

\author{L. Doyennette}
\affiliation{Laboratoire Charles Coulomb, Université de Montpellier, CNRS, 34095 Montpellier, France}

\author{B. Alloing}
\affiliation{University Côte d’Azur, CNRS, CRHEA, rue B. Gregory, 06560 Valbonne, France}

\author{E.~Cambril}
\affiliation{Centre de Nanosciences et de Nanotechnologies, CNRS, Universit\'e Paris-Saclay, France}

\author{S.~Bouchoule}
\affiliation{Centre de Nanosciences et de Nanotechnologies, CNRS, Universit\'e Paris-Saclay, France}

\author{T.~Ackemann}
\affiliation{University of Strathclyde, Glasgow, UK}

\author{J.~Z\'u\~niga-P\'erez}
\affiliation{University Côte d’Azur, CNRS, CRHEA, rue B. Gregory, 06560 Valbonne, France}
\affiliation{MajuLab, International Research Laboratory IRL 3654, CNRS, Université Côte d’Azur, Sorbonne Université, National University of Singapore, Nanyang Technological University, Singapore, Singapore}
\affiliation{School of Electrical and Electronic Engineering, Nanyang Technological University, Singapore}

\author{D. Solnyshkov}
\affiliation{Institut Pascal, PHOTON-N2, Universit\'e Clermont Auvergne, CNRS, Clermont INP,  F-63000 Clermont-Ferrand, France}
\affiliation{Institut Universitaire de France (IUF), 75231 Paris, France}

\author{G. Malpuech}
\email{guillaume.malpuech@uca.fr}
\affiliation{Institut Pascal, PHOTON-N2, Universit\'e Clermont Auvergne, CNRS, Clermont INP,  F-63000 Clermont-Ferrand, France}

\author{T. Guillet}
\email{Thierry.Guillet@umontpellier.fr}
\affiliation{Laboratoire Charles Coulomb, Université de Montpellier, CNRS, 34095 Montpellier, France}

\begin{abstract}
We study the emergence of multi-soliton regimes in 1D ridge polariton waveguides of two different lengths. We show that by varying the position of the gain, which in out-of-equilibrium polariton systems is provided by the pumping laser and its associated excitonic reservoir, it is possible to tune the regime of soliton formation between single and multiple solitons. This soliton dynamics can be quantitatively reproduced by solving the Gross-Pitaevskii equations of the coupled exciton-photon system, which show that the  soliton splitting mechanism is governed by the exciton reservoir dynamics. 
\end{abstract}
\maketitle

\section{Introduction}

Mode-locking is useful in achieving  frequency combs for high-precision metrology~\cite{jankowski2024}. It involves soliton pulses propagating inside the optical cavity, generated by the interplay between interactions and dissipation ~\cite{picholle1991observation}. 
Single and multiple cavity solitons have been widely investigated in mode-locked lasers based on fibers and VECSELs~\cite{Grelu_2012}. Multiple solitons can be either synchronous (also called soliton molecules~\cite{grelu2002phase,Stratmann2005}) or asynchronous~\cite{noske1994dual,zhao2011switchable}. Both types have their applications~\cite{hu2017asynchronous}: synchronous solitons can be used for Raman spectroscopy~\cite{ganikhanov2006broadly}, terahertz generation~\cite{Majkic2014} and timing distribution, while asynchronous soliton applications include asynchronous optical sampling~\cite{zhao2012fast}, terahertz frequency measurement~\cite{hu2017measurement}, and broadband optical and terahertz spectroscopy~\cite{zhao2016picometer}. Multiple coexisting asynchronous solitons were observed experimentally in many works with fiber lasers, such as~\cite{Malica_2018, wang2020real}.

An alternative platform for fabrication of mode-locked lasers is represented by microcavities operating in the strong light-matter coupling regime. Compared to VECSELs, with which they share many similarities, the strong coupling of excitons and photons giving rise to exciton-polaritons provides enhanced non-linear interactions together with dissipation, which are both required for soliton formation~\cite{Aranson2002,Bobrovska2015,lagoudakis2017polariton}.

The possibility of achieving mode-locked lasing with polaritons of various types was first studied theoretically in~\cite{kim2011theory,bagayev2017polariton}.
Soon after, solitons with exciton-polaritons were successfully observed in planar cavities~\cite{sich2012observation}, where e-beam lithography and etching allow engineering various photonic structures~\cite{tanese2013polariton,Goblot2019}, including 1D topological lattices displaying gap solitons~\cite{pernet2022gap}. 
In parallel to vertical microcavities containing a large number of distinct material layers, polariton waveguides have attracted a lot of attention for the simplicity of their fabrication and the long lifetimes of the associated guided photonic and polaritonic modes~\cite{beggs2005waveguide,solnyshkov2011polariton,walker2013exciton,solnyshkov2014optical,mechin2025time},
allowing one to observe nonlinear amplification and polariton lasing~\cite{jamadi2018edge,walker2019spatiotemporal,ciers2020polariton,suarez2020electrically,suarez2021enhancement,di2021ultrafast}. Bright and dark solitons have already been observed in polariton waveguides~\cite{Walker_2015,walker2017dark}.

The research on exciton-polaritons in waveguide cavities lead recently to the first demonstration of a mode-locked polariton ridge laser based on a 60~$\mu$m-long GaN in-plane Fabry-Perot cavity~\cite{optica2024}, with an energy per pulse as low as 5~fJ. 
It was shown that the pumping spot size, which defines the gain region and can cover in polariton ridge lasers just a fraction of the ridge length (potentially much smaller than the total length), does not affect significantly the formation of a soliton.


In this work, we study the behavior of polariton mode-locked lasing in two waveguide cavities displaying very different cavity lengths (compared to the dispersion and nonlinear lengths), as a function of the position of a pump whose size is systematically much shorter than the cavity length. A crucial effect arises whenever, for a constant ridge fraction covered by the pump, the gain region position is swept along the ridge length. In particular, we demonstrate a continuous conversion of a single soliton into a pair of asynchronous solitons, which provide dual lasing at close energies, when the position of the pump is shifted away from the cavity center. In even longer in-plane cavities, we observe that the mode-locking instability exhibits a periodicity of approximately 20~$\mu$m. The measurements of the photon statistics confirm the simultaneous existence of two solitons at different energies within each ns-long excitation pulse.

\section{Multiple soliton formation}

\begin{figure}
    \centering
    \includegraphics[width=1.0\linewidth]{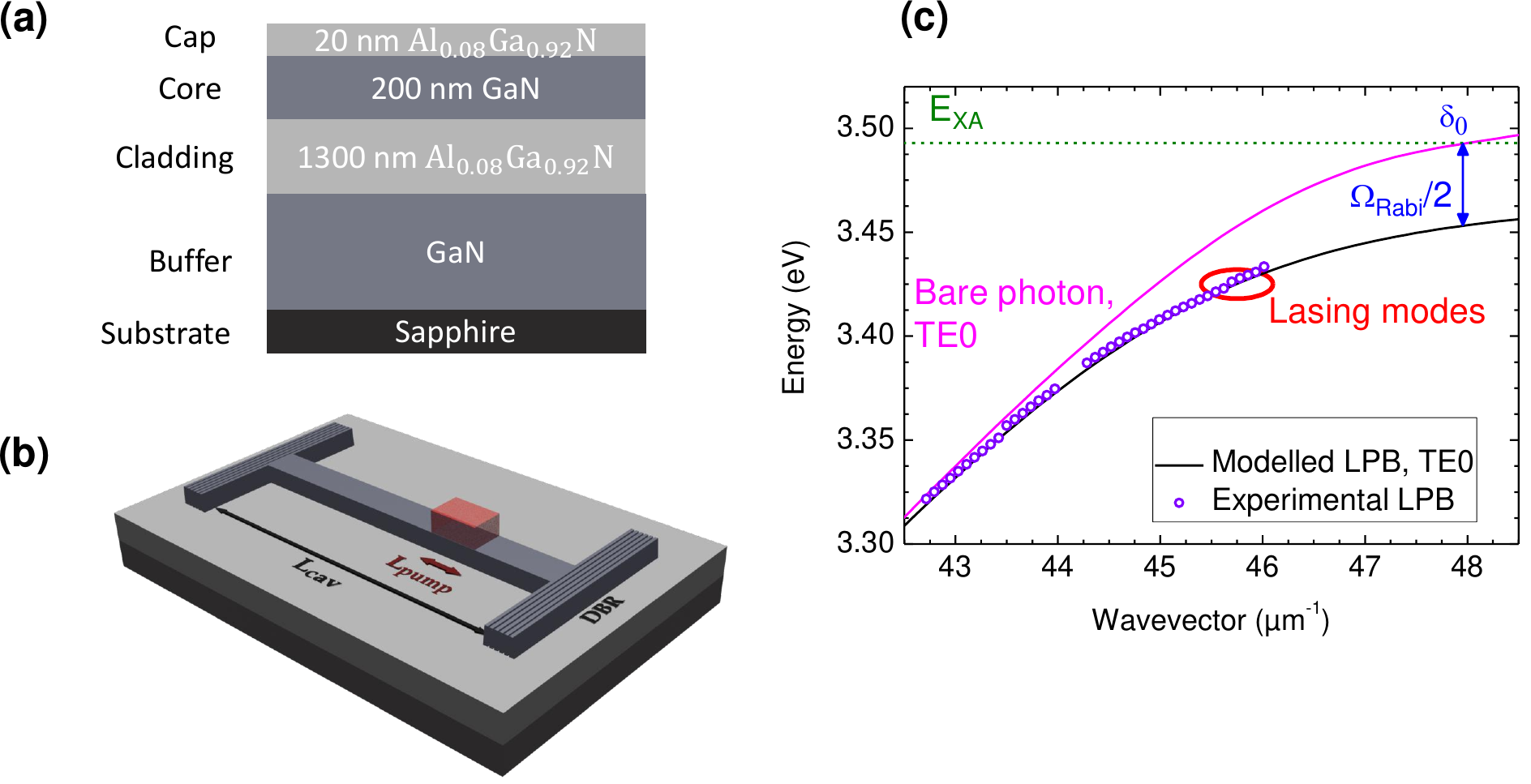}
    \caption{The GaN polariton waveguide cavity. (a) Scheme of the vertical cross-section of the waveguide, (b) 3D sketch of the Fabry-Perot ridge cavity formed by the 2 DBRs, with the pumping region shown in red; (c) Dispersion of the polariton branch at 150~K (violet circles) extracted from the FSR measurements, compared with the computed dispersion of the bare photon (magenta). The exciton resonance is shown as a green dashed line. The lasing modes are delimited by the red circle.}
    \label{fig1}
\end{figure}

The GaN waveguide slab consists of a bottom 1.3~$\mu$m-thick AlGaN cladding with an 8\% Al composition, a 200~nm-thick GaN waveguide core, and a top 20~nm cap of AlGaN with the same composition as the bottom cladding. The whole heterostructure was grown by metalorganic-vapour phase epitaxy on a 3~$\mu$m-thick GaN template on sapphire displaying a dislocation density of about $3\times 10^8$~cm$^{-2}$. The cross-section of the sample is shown in Fig.~\ref{fig1}(a). Using electron beam lithography and inductive plasma etching we obtained a ridge waveguide structure with DBRs on its edges, as shown in Fig.~\ref{fig1}(b), which define the in-plane Fabry-Perot cavities. The excitonic resonances in the core GaN layer are strongly coupled with the guided TE$_0$ photonic mode, as confirmed by the measured dispersion of the guided polaritons, shown in Fig.~\ref{fig1}(c). This dispersion is obtained by integrating the free spectral range (FSR)~\cite{optica2024}. However, it has also been measured directly on a similar waveguide structure with a grating~\cite{mechin2025time}.

\begin{figure}
    \centering
    \includegraphics[width=0.9\linewidth]{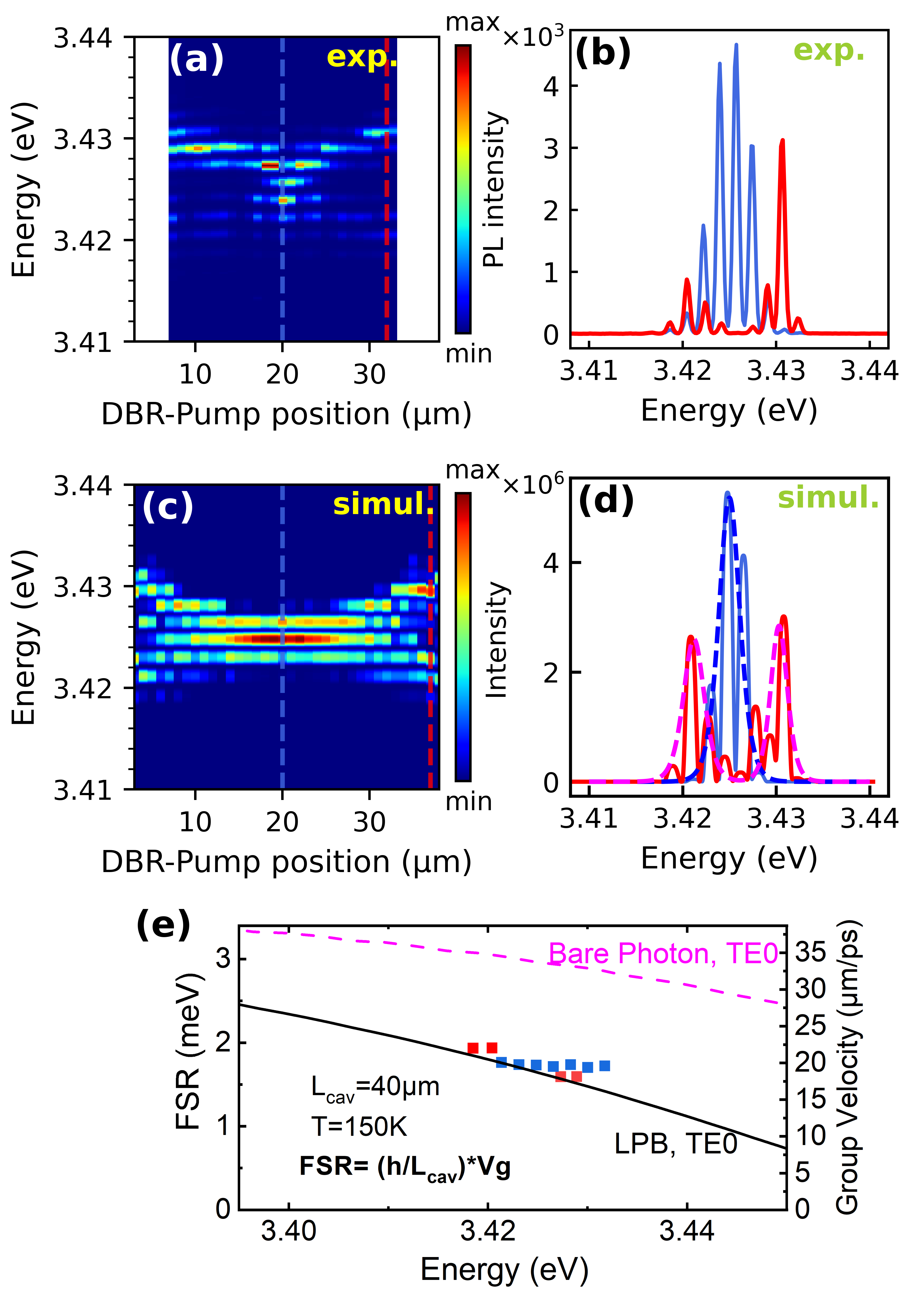}
    \caption{The transition between one- and two-soliton regimes. 
   (a) Spectrally-resolved intensity of the emission measured in the experiment for different pump positions.  (b) Spectra for peculiar distinct regimes (achieved by pumping at the positions indicated by the blue and red dashed lines in (a)) illustrating the presence of one and two bright temporal solitons (in blue and red, respectively). The middle row (c-d) shows the corresponding simulations results. The dashed curves in (d) are for the $\sech$ fit. (e) Free spectral range (FSR) calculated for the experimentally measured spectra in (b) for the one- (in blue) and two-soliton (in red) cases. Two distinct group velocities are measured in the latter regime, indicating formation of two asynchronous temporal solitons. The pink dashed and black solid curves mark the bare photon and polariton dispersions (see Fig.~\ref{fig1}) and are added as a guide for the eye. }
    \label{fig2}
\end{figure}

We now focus on the mode-locking and soliton formation in this ridge waveguide.
Figure~\ref{fig2} presents the case of a $40~\mu$m-long cavity pumped with a line-shaped spot (4~ns pulses at 355~nm at a repetition rate of 7~kHz, resonant with the exciton energy in GaN) with a spatial extension of $6.5~\mu$m, at $T = 150$~K.
Panel~(a) shows the spectrally-resolved emission intensity collected at one of the DBRs (i.e. spatially filtered) as a function of the pump position across the laser ridge. Clearly, changing the position of the pump strongly changes the spectral distribution of the emission intensity. Two spectra measured for two different pump positions, corresponding to the dashed blue and red lines in Fig.~\ref{fig2}(a), are shown in Fig.~\ref{fig2}(b). The spectrum plotted in blue, corresponding to a pump positioned on the cavity center, demonstrates a single soliton: the emission  exhibits a characteristic hyperbolic secant envelope structured by the Fabry-Perot cavity modes. Note that the same spectrum is obtained when pumping the whole cavity length, as in ref. \cite{optica2024} (not shown). However, moving the pump closer to one of the Bragg reflectors (red dashed line in Fig.~\ref{fig2}(a)) induces an instability, which results in switching to a multiple-soliton regime (red curve in Fig.~\ref{fig2}(b)). This spectrum demonstrates the presence of two distinct asynchronous solitons lying at different energies. In order to quantitatively evidence the latter fact, we plot in Fig.~\ref{fig2}(e) the FSRs extracted from the corresponding spectra for the one and 2-soliton cases (in blue and red squares, respectively). In each case, we observe mode locking: for a single soliton (blue squares) it involves all modes of this soliton, while for two solitons (red squares) it occurs independently for each of them, confirming their asynchronous nature. The two solitons  possess two well-defined group velocities ($15.4~\mu$m/ps and $18.7~\mu$m/ps), which are in turn distinct from the one of the single-soliton case.

In order to theoretically investigate the dynamics of bright solitons, we perform numerical simulations using coupled 1D Gross-Pitaevskii equations for photons $\varphi$, excitons $\chi$ and the excitonic reservoir $n_R$, which can be written as follows:
\begin{widetext}
\begin{eqnarray}\label{gpe3eq}
    &&\hspace{0.5cm} i\hbar\frac{\partial\varphi}{\partial t}= [\hat{T}-\frac{i\hbar}{2\tau_{Ph}} ]\varphi-\Omega_R\chi ,\nonumber\\
   && \hspace{0.5cm} i\hbar\frac{\partial\chi}{\partial t}= -\Omega_R\varphi+g\vert\chi\vert^2\chi+V_{pot}\chi-\frac{i\hbar}{2\tau_{Ex}}\chi+\frac{i\hbar}{2}R\hat{\Gamma }(\chi)n_R+2g_{Res}n_R \hat{\Gamma}(\chi)+\xi_{noise},\nonumber\\
  &&  \hspace{0.5cm} \frac{\partial n_R}{\partial t} =P-R\vert\hat{\Gamma}(\chi)\vert^2 n_R-\frac{1}{2\tau_{Res}} n_R.
\end{eqnarray}
\end{widetext}
Here $\Omega_R=80$ meV indicates the Rabi splitting, $g$ and $g_{Res}$ stand for the polariton interaction constants, $\tau_{Ph}$, $\tau_{Ex}$, $\tau_{Res}$ are the photon, the exciton and the exciton reservoir lifetimes, respectively. The reservoir-condensate scattering rate, $R$, is responsible for the stimulated scattering of polaritons along their dispersion. 
$\hat{T}$ stands for the kinetic energy and $\hat{\Gamma}(\chi)$ describes the polariton gain, which plays a key role in capturing the effects of the polariton energy relaxation. 

The numerical simulations were performed on a 1D grid
of 1280 points resulting in a spatial resolution of about 0.03~$\mu$m.
The integration is done using the 3-step Adams-Bashforth method with a time step of $10^{-4}$~ps.  We use the Fourier transform~$\mathfrak{F}$ to treat the kinetic energy and gain terms. The gain profile in the momentum space is taken to select the states of the polariton dispersion centered at the wave vector $k_0$, $G(k)=\exp\{-(k-k_0)^2/(2 w^2)\}$.  On the other hand, in real space the gain is shaped by the pump profile $p(x)$, so that $\hat{\Gamma}(\chi)=\mathfrak{F}[G(k)\mathfrak{F}(\chi(x))]p(x)$. We emphasize that despite the fact that $G(k)$ is a symmetric function, the resulting net polariton gain profile becomes strongly asymmetric due to the dispersion shape.

The results of the simulations are shown in Fig.~\ref{fig2}(c,d). Panel~(c) shows the obtained intensity of emission from the edge of the sample 
as a function of the pump position $x$ and energy $E$, as in the experiment. At the same time, panel (d) shows two characteristic examples of spectra simulated for two pump positions indicated by blue and red dashed vertical lines in (c), equivalently to the experimental ones shown in Fig.~\ref{fig2}(b).
In Fig.~\ref{fig2}(c) we observe a continuous transition from the single-soliton to the multi-soliton regime by changing the pumping position, which modifies the interaction of the solitons with the reservoir, as discussed below, and which reproduces the experimental observations.

\begin{figure}
    \centering
    \includegraphics[width=0.95\linewidth]{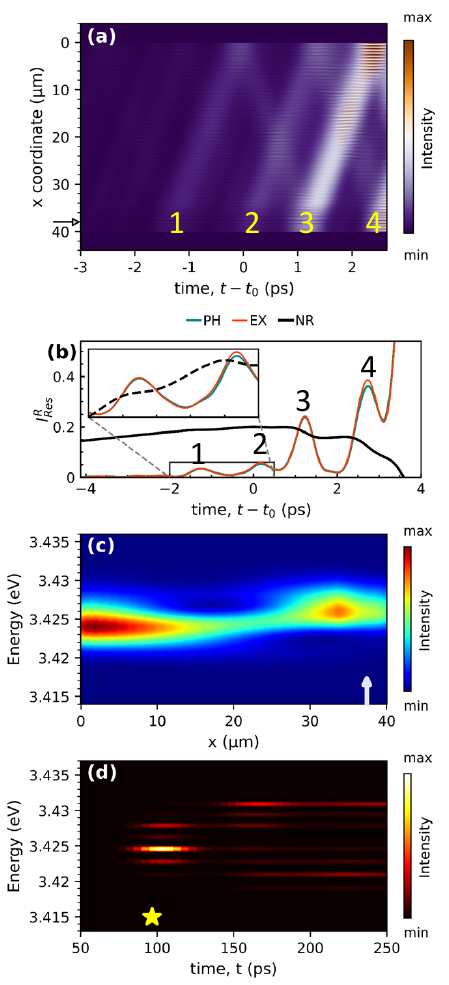}
    \caption{Two-soliton regime. (a) Real space intensity. The arrow on the left indicates the position of the pump center. Numbers mark reamplifications of the solitons via the reservoir. (b) Rescaled intensities of the photon and exciton fractions together with the population of the excitonic reservoir (black curve) integrated over the pump region. Inset shows a zoom, where the dashed black curve corresponds to the rescaled black one from the main subfigure.  (c) Emission intensity as function of position and energy for the case where more than one soliton is present
    in the cavity with distinct group velocities, and integrated for a time window (centered at $t_0$) chosen to be shorter than roundtrip time, $t_w=0.8$~ps. The white arrow indicates the location of the pump center. (d) The time-resolved spectral intensity  $t_w\approx 30$~ps showing formation of two asynchronous solitons. The yellow star indicates $t_0$.}
    \label{fig6}
\end{figure}

Qualitatively, the possibility of having multiple solitons is determined by the ratio of characteristic times: the soliton round-trip time (from the reservoir to the farthest mirror) and the mode refilling time. 
Figure~\ref{fig6} demonstrates multiple aspects of this process.
Fig.~\ref{fig6}a shows the real space intensity of the photonic fraction for the case when the pump is located close to one of the DBRs (as indicated by the black arrow). This configuration is the same as marked by the red dashed line in Fig.~\ref{fig2}(a). Further, Fig.~\ref{fig6}(b) shows the populations of photons (cyan), excitons (orange) and the exciton reservoir (black) as functions of time, integrated over the pumping region and rescaled for presentation purposes.
In this case, the characteristic soliton round-trip time, given by $(L+2x_R)/v_g$ and determined by the group velocity, is longer than the mode refilling time, determined by the spontaneous scattering time from the reservoir $\tau_m\sim\xi^{-1}$ to the photonic modes. Thus, after the generation of the first soliton (marked by "1" in Fig.~\ref{fig6}(a-b)) and before its return to the pump location, the conditions for the generation of the second soliton (marked "2") become satisfied too. The generation of the soliton "2" drains the mode, which has however enough time to be replenished once again before the soliton "1" comes back. It is therefore reamplified (marked "3"). The reamplification of the soliton "2" gives the soliton "4".  The intensity of the solitons increases with each new round, until the soliton becomes sufficiently populated to drain the reservoir completely in the last pass, which occurs at around $t_0+4$~ps. Here and below $t_0$ marks the time moment which corresponds to the center of the time window for the Fourier transform for  Fig.~\ref{fig6}(b-c), that is, when the two solitons "1" and "2" are visibly separated in space over the cavity.  

The peaks of the emission intensity are followed by the drops in the reservoir population, as highlighted in the inset of Fig.~\ref{fig6}(b). Importantly, the net polariton gain is not the same at the moments when the two solitons "1" and "2" are generated. As a result, the two of them can have slightly different average energies and, therefore, group velocities. This is confirmed by performing Fourier transform centered at time $t_0$ (with a sufficiently small time window $t_w=0.8$~ps), as presented in Fig.~\ref{fig6}(c), where spatially-resolved spectral intensity is presented~\cite{rq1}. One can see that the two solitons with different energies are separated spatially within the cavity, which is also visible in Fig.~\ref{fig6}(a). The white arrow indicates the pump position.

Finally, Fig.~\ref{fig6}(d) shows the spectral intensity resolved in time on a relatively large time scale and with a large (time) averaging. In this case the time window of the Fourier transform is taken to be much longer than the soliton roundtrip period in order to smooth out the spatial dependence: $t_w=30$~ps. From the spectrum close to time  $t_0$, indicated by the yellow star, we can conclude that two coherent solitons appear during the condensation: this is manifested as a suppression of one of the central peaks in the spectrum by their interference. 

Subsequent long (multiple round-trip) propagation of solitons in the absence of gain during the long reservoir replenishment after its strong draining by the soliton "4"  results in the loss of the relative coherence of the two coexisting solitons. Here, another two  characteristic times are involved: the loss of coherence occurs on the time scale $2L/\Delta v_g \approx 25$~ps, when the two solitons become spatially separated and cannot interact any more. This timescale needs to be shorter than the so-called large-amplitude reservoir relaxation oscillations,~\cite{Opala2018} determined by the pumping $P$ and the amplitude of the "active" part of the reservoir $\Delta n_R$.
Similar effects occur with slow saturable absorbers~\cite{LSAinbook,Kurtner1998}, which are known for providing passive mode-locking in lasers~\cite{Kurtner1998, Kartner1995, Grelu_2012}. Such mechanisms of mode competition cause, e.g., temporal mode switching, recently observed for polariton condensates~\cite{Urbonas2024}.

\begin{figure}
    \centering
    \includegraphics[width=0.95\linewidth]{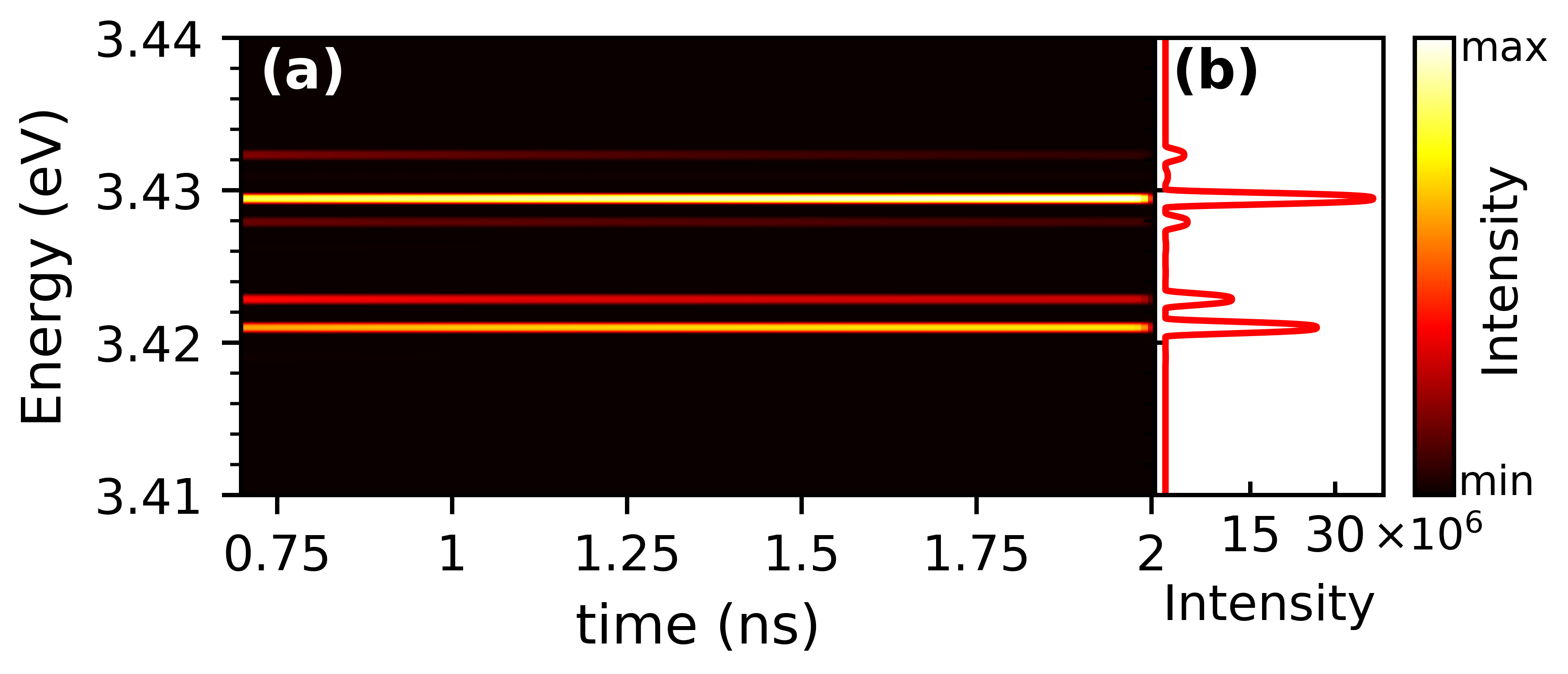}
    \caption{(a) Time-resolved spectral intensity for the two-soliton regime ($t_w=30$ ps). (b) the corresponding time-integrated spectrum.}
    \label{fig5}
\end{figure}

In order to show the stability of the two-soliton regime in stationary conditions, we calculate and plot in Fig.~\ref{fig5} the time-resolved spectra obtained by implementing sliding-window ($t_w=30$~ps) Fourier transform, for the situation where the pump is close to one of the mirrors. One can see the simultaneous presence of two temporal solitons in the cavity over sufficiently long times. This evidences their stability and the robustness of the detection. 

\begin{figure}
    \centering
    \includegraphics[width=1.0\linewidth]{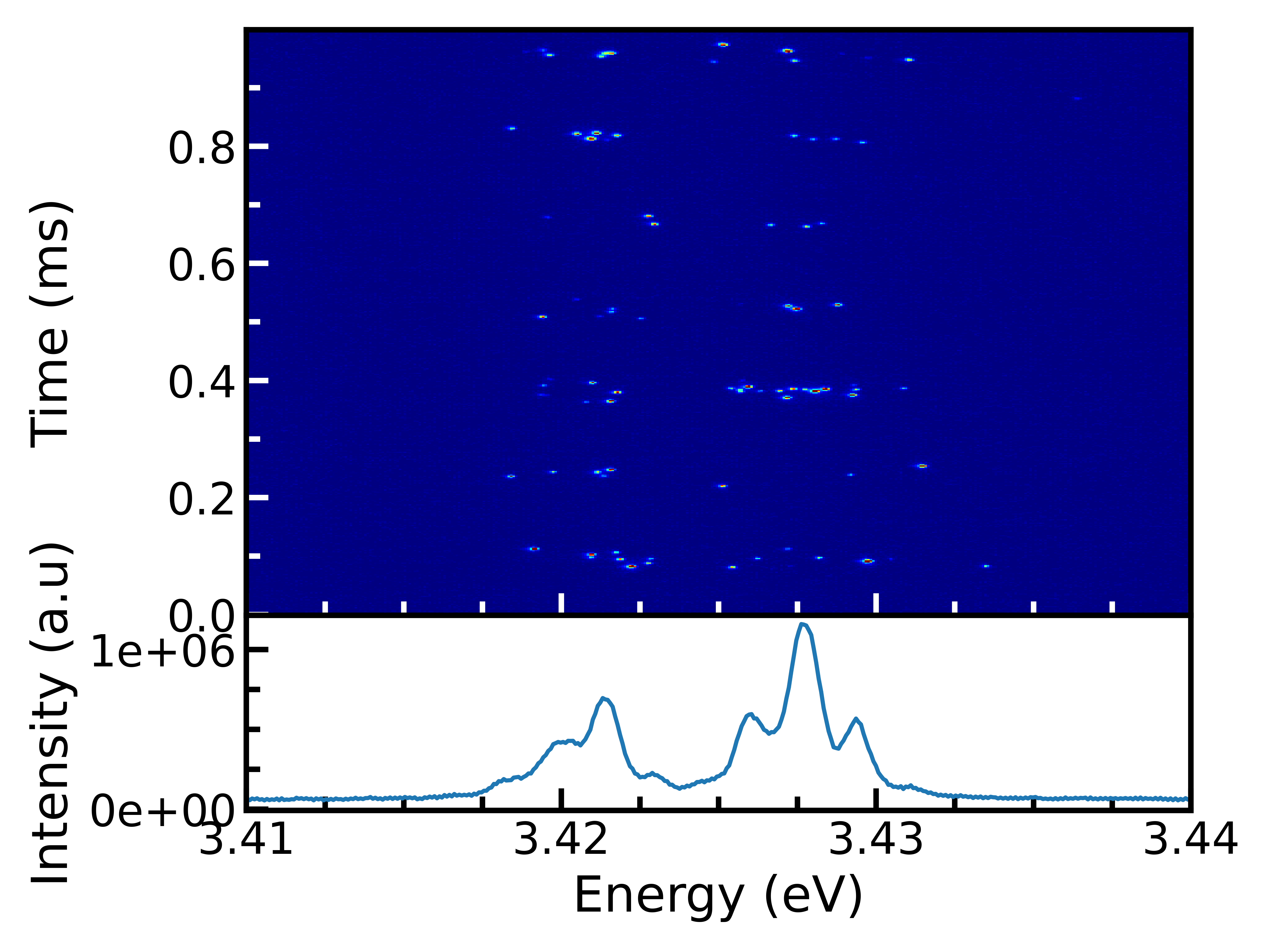}

    \caption{Simultaneous existence of two solitons. Number of photons collected from multiple repetitions of the experiment as a function of the energy and time. Lower panel shows integrated spectrum.}
    \label{fig4}
\end{figure}

In order to demonstrate that the solitons observed experimentally in the integrated spectra exist simultaneously in time, and are not just due to pulse-to-pulse random fluctuations of the experimental conditions (e.g. of the ns-long pump pulses), we present experimental results of single-shot measurements providing the number of photons as a function of wavelength for multiple repetitions of the pulse in the multi-soliton regime. The polariton laser being pumped by 4ns-long pulses every $140~\mu$s, the emission is temporally resolved with a streak-camera coupled to the spectrometer. A single frame (showing multiple pumping pulses without any averaging) is shown in Fig.~\ref{fig4} (top):  while the overall number of detected photons is relatively low and the temporal resolution does not allow to make any conclusions about the dynamics within a single pulse, we can nevertheless conclude that the two solitons at different energies exist for each pulse. The bottom panel of the same figure shows the signal integrated over all measured frames, forming the two-soliton spectrum similar to Fig.~\ref{fig2}(b).

\begin{figure}
    \centering
    \includegraphics[width=1.0\linewidth]{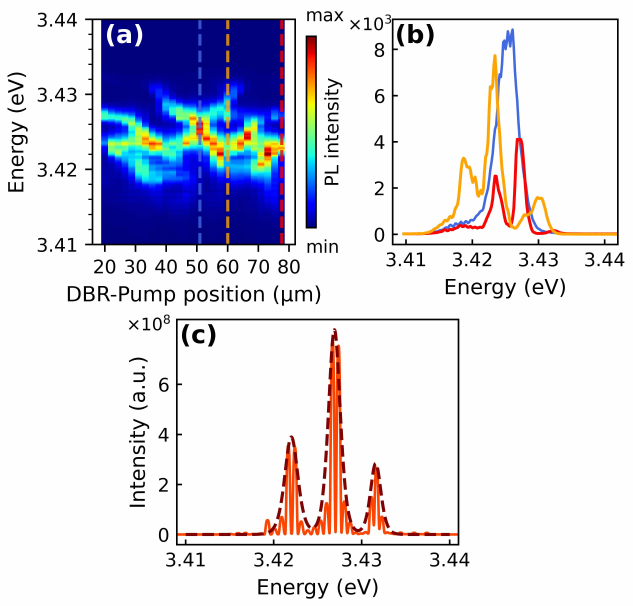}
    \caption{Soliton formation for larger sample of $100~\mu$m. (a) Intensity of emission as a function of pump position and energy. (b) Spectra for highlighted pump position demonstrating one (blue), two (red) and three-soliton (orange) cases. (c) Three-soliton spectra (light red curve) with the fit (dashed curve) obtained from numerical calculations, analogously to the orange curve in (b).}
    \label{fig3}
\end{figure}

Interestingly, more than 2 coexisting solitons can be generated in longer cavities. 
This can be qualitatively understood by comparing the cavity length to the typical dispersion length ($L_D = 25~\mu$m) (related to the variation of the group velocity) and to the nonlinear length ($L_{NL} \approx 50~\mu$m) estimated previously in these lasers~\cite{optica2024}. This comparison gives an indication of how many incoherent solitons can be simultaneously contained within the cavity. The results of experimental measurements in a 100~$\mu$m cavity with a 20~$\mu$m pump (i.e. a ratio similar to the one between the pump length and the cavity length) are shown in Fig.~\ref{fig3}. Again, a single soliton forms when the cavity is pumped precisely at the center. The transition between one and multiple-soliton regimes occurs periodically on the scale of about 20~$\mu$m. Three solitons can be observed in the spectrum for a given pump position, see Fig.~\ref{fig3}(b) (orange curve). Three simultaneous solitons are also observed in numerical simulations, see Fig.~\ref{fig3}(c), for the case when the pumping spot is  shifted from the center of the cavity. The overall pattern in Fig.~\ref{fig3}(a) is symmetrical with respect to the cavity center, and is identical when the same experiment is performed on a different 100~$\mu$m-long cavity on the same sample. This demonstrates that photonic disorder and inhomogeneities in the reservoir density are not responsible for the formation of multiple solitons, which is solely determined by the interplay between the reservoir dynamics and the nonlinear dynamics.

\section{conclusions}

After the first observation of polariton mode-locked lasing in waveguide cavities, which allows the generation of narrow frequency combs, we show here that thanks to the crucial capacity of polariton lasers to operate under localized pumping (i.e. covering only a small fraction of the laser ridge length), it is possible to use the position of the pump as a control knob to tune the mode-locking regime from the single-soliton to the multiple asynchronous solitons regime. We demonstrate both experimentally and numerically the formation of two and three solitons. Their appearance is solely controlled by the position of the pump with respect to the mirrors establishing the lasing feedback, with a pattern that repeats with a period of approximately 20~$\mu$m. Single-shot measurements confirm the simultaneous existence in time of multiple solitons, supporting the claim of dual lasing.

\begin{acknowledgments}
This work was supported by European Union's Horizon 2020 program, through a FET Open research and innovation action under the grant agreement No. 964770 (TopoLight), and by the European Union EIC Pathfinder Open project “Polariton Neuromorphic Accelerator” (PolArt, Id: 101130304). Additional support was provided by the ANR Labex GaNext (ANR-11-LABX-0014), the ANR program "Investissements d'Avenir" through the IDEX-ISITE initiative 16-IDEX-0001 (CAP 20-25), the ANR project MoirePlusPlus, the ANR project "NEWAVE" (ANR-21-CE24-0019), and the R\'egion Occitanie. C2N is a member of RENATECH-CNRS, the French national network of large micro-nanofacilities.
\end{acknowledgments}

\appendix
\section{Experimental setup}
The sample is mounted in a cold-finger cryostat (\textit{Oxford Microstat Hires2}) and maintained at a temperature of 150~K. It is excited by a pulsed laser pump (355~nm at a repetition rate of 7~kHz, \textit{Cobolt Tor}). The shape of the pump beam is controlled using a variable rectangular slit and a cylindrical lens placed before the microscope objective, which homogeneously expands the pump beam in one direction. The rectangular slit opening controls the pump length, which can be varied from 6.5 to hundreds of micrometers, while the position of the pump on the cavity is controlled by piezoelectric actuators.

The microscope objective (\textit{Mitutoyo M plan APO NUV}) focuses the pump onto the sample and collects the resulting emission from the cavity. The cavity image is projected onto the entrance slit of the spectrometer (\textit{Horiba Jobin Yvon iHR550}). A charge-coupled device (\textit{Andor Newton CCD}) and a streak camera (\textit{Hamamatsu C109010}) are mounted at the spectrometer output ports. Using a diffractive grating of 1200~grooves/mm, the system enables the reconstruction of real-space or time-resolved photoluminescence images, with the CCD or the streak camera, respectively.

\bibliography{biblio,u_biblio_2}

\end{document}